\documentclass[ letterpaper, 10pt, conference]{ieeeconf}

\pdfoutput=1
\usepackage{graphicx}
\usepackage{float}
\usepackage{amsfonts}
\usepackage{amsmath}
\usepackage{amssymb,latexsym}
\usepackage{cite}
\usepackage{bm}
\usepackage{mathrsfs}
\usepackage{subfig}
\usepackage{wrapfig}
\newtheorem{definition}{Definition} 
\newtheorem{lem}{Lemma}

\newtheorem{thm}{Theorem}
\newtheorem{prob}{Problem}
\newtheorem{assum}{Assumption}
\usepackage[font=small]{caption}
\usepackage{xcolor}
\newcommand{\ol}{\overline}
\usepackage{booktabs}
\IEEEoverridecommandlockouts
\begin{document}
	
	\title{Safety Verification of Output Feedback Controllers for Nonlinear Systems}
	\author{Kendra Lesser$^{1}$ and Alessandro Abate$^{1}$%
		\thanks{}%
		\thanks{$^{1}$Department of Computer Science, University of Oxford, UK, {\tt\small \{kendra.lesser, alessandro.abate\}@cs.ox.ac.uk}
		}
	}
	\maketitle
		
	\begin{abstract}

		A high-gain observer is used for a class of feedback linearisable nonlinear systems to synthesize safety-preserving controllers over the observer output.  A bound on the distance between trajectories under state and output feedback is derived, and shown to converge to zero as a function of the gain parameter of an observer.  We can therefore recover safety properties under output feedback and control saturation constraints by synthesizing a controller as if the full state were available.  We specifically design feedback linearising controllers that satisfy certain properties, such as stability, and then construct the associated maximal safety-invariant set, namely the largest set of all initial states that are guaranteed to produce safe trajectories over a given (possibly infinite) time horizon.
	\end{abstract}
	
	\section{Introduction}
	
	Verification is an increasingly important aspect of control system design. In particular for safety-critical systems, such as aircraft, satellites, and autonomous vehicles, it is crucial to certify that the controllers designed for these systems will not lead to costly failures, namely that the state of the system will not reach unsafe regions of the state space.  Model-based verification is gaining traction as a tool for assessing safety of control systems \cite{TabuadaBook}, but is only as accurate as the model being used. An often overlooked model feature is the unavailability of the full state of the system for feedback control.  In such cases a controller may be designed as if the full state were available, and then a state estimate is used in place of the actual state.  In doing so, any properties the controller is designed for (linearisation, stability, safety) are no longer guaranteed.
	
	For nonlinear systems, it is difficult to design feedback controllers to achieve desired performance, even when the full state is known.  Linear systems, on the other hand, are well-studied with an array of tools available for controller design.  If possible, it is therefore desirable to feedback linearise a nonlinear system to then apply techniques from linear control theory \cite{sastry}.  This technique, however, requires full knowledge of the state and no uncertainty in the model.
	
	 When the state is unknown, or known only partially, an observer can be used to estimate it.  For systems transformable to normal form \cite{sastry}, a nonlinear high-gain observer (HGO) can rapidly reconstruct the actual state because the error dynamics quickly stabilize \cite{Khalil_IJRNC14}.  If the HGO output is used in place of the actual state in a feedback controller, the output feedback trajectory approximately recovers the trajectory that would arise under state feedback, after a brief transient period in which the estimation errors are large \cite{Khalil_TAC99, Teel_SJCO95}.  The implication for feedback linearisable systems is that under certain conditions, linearisation (approximately) still holds under output feedback, and controller synthesis techniques for linear systems still apply \cite{Khalil_TAC08}.  These results, however, typically are used only for stabilization and reference tracking objectives (as in \cite{Khalil_IJRNC14, Khalil_TAC99, Teel_SJCO95, Khalil_TAC08}), and have not been utilized in a verification context.
	 
	 Alternately, verification and controller synthesis with safety objectives are well-studied for fully observable systems.  A common approach for nonlinear systems is to use an optimal control formulation that relies on Hamilton-Jacobi-Bellman equations and discretization-based approximate solutions, to find the maximal set of initial conditions that lead to trajectories that do not violate safety constraints \cite{Tomlin98_TAC}, \cite{Mitchell2005_TAC}, \cite{Marg1}.  Feedback linearising controllers are synthesized in \cite{Oishi_CDC06}, using the above formulation to ensure that safety specifications are satisfied and that the controller does not saturate.  
	 
	 Controller synthesis for \emph{partially observable} systems that are guaranteed to obey safety constraints (i.e. correct-by-design control) is much less studied.  For stochastic systems, \cite{Lesser2014_Aut}, \cite{Ding2013} provide an optimal control formulation to synthesize safety maximizing controllers, which use as input a probability distribution that captures any knowledge about the state of the system.  An output feedback controller linked to an observer is synthesized and proven to satisfy safety specifications for linear stochastic systems in \cite{Lesser_ADHS15} and for linear deterministic systems in \cite{Sofie_CDC15}.  Correct-by-design output feedback controllers are synthesized for a class of switched linear systems in \cite{Mickelin_ACC14}.     
	 
	 This work, to the best of our knowledge, is the first to synthesize controllers with safety guarantees for a class of \emph{nonlinear} systems.  We consider full state feedback linearisable systems in normal form, and construct a high-gain observer in order to recover the desired trajectory using output feedback.  \emph{Our contributions are a) the derivation of explicit bounds on the distance between trajectories under state feedback and trajectories under output feedback using an HGO, and b) the synthesis of feedback linearising controllers that satisfy certain specifications (such as stability, trajectory tracking, etc.) that are also provably safe and do not violate any control saturation constraints.} We achieve this by utilizing the bound we derive between trajectories to design controllers as if the state were available for feedback, and by using a more conservative safety constraint that is a function of that bound.  
	
%
%
%
%
%
%
%
%
%
%
%
%
%
%
%
%
	
	\section{Background}

	\subsection{Feedback Linearisable Systems}
	
	We consider a nonlinear single-input, single-output system in normal form \cite{sastry} that is full state feedback linearizable: 
	\begin{equation}
	\label{eq:dynamics}
	\begin{aligned}
	\dot{x} &= Ax + B[b(x) + a(x)u] \\
	y &= Cx, 
	\end{aligned}
	\end{equation}
	with 
	\begin{align*}
	A &= \begin{bmatrix} 0 & 1 & 0 & \cdots & 0 \\ 0 & 0 & 1 & \cdots & 0 \\ \vdots & \vdots & \ddots & \ddots & \vdots \\ \vdots & \vdots & \cdots & 0 & 1 \\ 0 & 0 & \cdots & \cdots & 0 \end{bmatrix} \in \mathbb{R}^{n\times n}, \hspace{3 mm} B = \begin{bmatrix} 0 \\ 0 \\ \vdots \\ 0 \\ 1\end{bmatrix} \in \mathbb{R}^{n}, \\
	 C &= \begin{bmatrix} 1 & 0 & \cdots & 0 \end{bmatrix} \in \mathbb{R}^n. 
	\end{align*}
	The state is $x\in\mathbb{R}^n$, 
	the output is $y\in\mathbb{R}$, 
	and the control input is $u\in\mathbb{R}$.  
	Any system with relative degree $n$ can be put into the form \eqref{eq:dynamics} \cite{sastry}. 
	For simplicity, in this work we assume that the system does not have zero dynamics, although an extension to this case is possible (see \cite{Khalil_TAC08}).  
	We also raise the following mild assumption on the model dynamics. 	
	
	\smallskip
		
		\begin{assum}\label{ass3}
			The functions $a(x)$ and $b(x)$ are locally Lipschitz continuous in $x$ for all $x\in\mathcal{A}\subseteq\mathbb{R}^n$, 
			locally bounded over $\mathcal{A}$, and $a(\cdot)\geq a_0>0$. \hfill$\square$
		\end{assum}
	
	\smallskip
		
	Given Assumption \ref{ass3}, if the state $x$ were available for feedback and there were no restrictions on the magnitude of the control input, 
	we could design a control law $u(t) = g(x(t))$, 
	\begin{equation}\label{eq:stateFB}
	g(x) = \frac{-b(x) + Kx+v}{a(x)}
	\end{equation} 
	to render the system linear, thus obtaining 
	\begin{equation}\label{eq:dynamicsFB}
	\dot{x} = (A + BK)x + Bv, 
	\end{equation}
	where $v$ is a reference signal. 
	A control signal $u(t)$, however, is in general subject to a saturation condition such as $|u(t)|\leq u_{\mathrm{max}}$, i.e. $u(t)\in\mathcal{U}_{\mathrm{safe}}$ with $\mathcal{U}_{\mathrm{safe}}=[-u_{\mathrm{max}}, u_{\mathrm{max}}]$.  
	This saturation condition is imposed as follows: 
	whenever $g(x)$ lies inside the set $\mathcal{U}_{\mathrm{safe}}$, it does not saturate, 
	but as soon as $g(x)\notin \mathcal{U}_{\mathrm{safe}}$, it takes the value $\pm u_{\mathrm{max}}$.  
	If we let $\ol{g}(x)$ represent the controller without saturation constraints, then the actual implemented controller is defined as follows: 
	\begin{equation}
	\label{eq:controller}
	g(x) = \begin{cases}
	-u_{\max}, &\text{if $\ol{g}(x)\leq -u_{\max}$} \\
	\ol{g}(x), &\text{if $-u_{\max}< \ol{g}(x)< u_{\max}$ } \\
	u_{\max}, &\text{if $\ol{g}(x) \geq u_{\max}$ }
	\end{cases}.
	\end{equation} 	
	Further, in view of \eqref{eq:dynamics} the state $x$ is not available to the controller as in \eqref{eq:stateFB}, and the output $y=Cx$ must be used instead.  
	We would therefore like to design a state estimate $\hat{x}\in \mathbb{R}^n$ that is as close to $x$ as possible, 
	in order to approximately feedback-linearise the system \eqref{eq:dynamics}. 


%
%
	
	\subsection{High Gain Observers}
	
	We use a high-gain observer (HGO) of the form 
	\begin{equation}
	\label{eq:observer}
	\dot{\hat{x}} = A\hat{x} + B[b(\hat{x}) + a(\hat{x})u] + H(\epsilon)(y - C\hat{x}) 
	\end{equation}
	to construct an estimate of the state $x$ of \eqref{eq:dynamics}. 
	High-gain observers have been used for nonlinear systems because of their robustness to model uncertainty and fast convergence to the true state \cite{Khalil_IJRNC14}. 
	The observer gain $H(\epsilon)$ is a function of a positive parameter $\epsilon$, 
	and takes the form 
	\begin{equation*}
	H(\epsilon) = \begin{bmatrix}\frac{\alpha_1}{\epsilon} & \frac{\alpha_2}{\epsilon^2}&\cdots & \frac{\alpha_n}{\epsilon^n} \end{bmatrix}, 
	\end{equation*}
	with the $\alpha_i$ selected such that the polynomial 
	\begin{equation*}
	s^n + \alpha_1 s^{n-1} + \alpha_2 s^{n-2} + \dotsc \alpha_{n-1}s + \alpha_n
	\end{equation*}
	is Hurwitz (roots in the open left half plane).  
	The selection of the $\alpha_i$ guarantees that the error dynamics, defined over the signal $(x - \hat{x})$, are stable.  
	By picking $\epsilon$ arbitrarily small, 
	the error dynamics converge to zero arbitrarily quickly.  
	However, HGOs suffer from an initial peaking phenomenon before rapidly converging to the true state (i.e. the state estimate at early stages can be very large compared to the actual state).  
	As such, if the controller $g(x)$ is replaced by $g(\hat{x})$, this leads to erroneous control inputs early on.
	One partial remedy that we employ is to saturate the controller to avoid large magnitude control inputs \cite{Khalil_IJRNC14}. 
	
	\subsection{Safety Verification Under Output Feedback}
	
	We are interested in constructing feedback linearizing controllers for nonlinear systems that simultaneously satisfy both state and control input constraints (for safety requirements and due to saturation, respectively), given that the state is not available for feedback.  More precisely, given a set $\mathcal{X}_{\mathrm{safe}}\subset\mathbb{R}^n$ that we would like the system trajectory $x(t)$ to remain within (and that lies within the set $\mathcal{A}$ over which Assumption \ref{ass3} is valid), we would like to find the set of \emph{all} initial states $x(0)$ (denoted as the safety-invariant set) which, under a class of feedback linearizing controllers, generate trajectories that remain within $\mathcal{X}_{\mathrm{safe}}$ and respect the control constraints $u(t)\in\mathcal{U}_{\mathrm{safe}}$ for all $0\leq t \leq T$, where $T\leq\infty$.   
	


	\smallskip				
				
	\begin{definition}\label{def:invariance}
		The safety-invariant set $\Delta$ is the set of all initial states $x(0)$ that, under a given feedback controller $g(x)$, produce trajectories $x(t)$ that remain within the safe set $\mathcal{X}_{\mathrm{safe}}$ without violating the control constraint $u(t)\in\mathcal{U}_{\mathrm{safe}}$ for all $0\leq t \leq T$, with $T\leq \infty$. \hfill$\square$
    \end{definition}
	
	\smallskip
	
	We additionally raise the following assumptions on $\mathcal{X}_{\mathrm{safe}}$ and the initial state $x(0)$ (the latter being non-trivial as it is required for subsequent results).  The notation $\partial \mathcal{X}$ is used to denote the boundary of set $\mathcal{X}$.
	\smallskip
	\begin{assum}\label{ass:xSafe}
		The set $\mathcal{X}_{\mathrm{safe}}$ is a compact set contained in $\mathcal{A}$ (from Assumption \ref{ass3}), and $x(0)\in\mathcal{X}_{\mathrm{safe}}\backslash \partial \mathcal{X}_{\mathrm{safe}}$ (so $x(0)$ must lie in the interior of $\mathcal{X}_{\mathrm{safe}}$). \hfill$\square$
	\end{assum}

	\smallskip		
	
	The linearising controller $g(x)$ may be designed to stabilize the system, track a given reference trajectory, etc., and the safety-invariant set tells us from which initial states $x(0)\in\mathcal{X}_{\mathrm{safe}}$ the trajectory must initialize to further satisfy given state and control constraints.
	We could as well introduce a parametrized controller $g(x,\beta)$, as in \cite{Oishi_CDC06}, to try and maximize the size of the safety-invariant set $\Delta$ over $\beta$. 
	If we use a parametrized controller, however, $\Delta$ will be maximal only with respect to the imposed structure on the controller that targets the other control objectives (e.g.,  linearization plus stability).  
	The interest in maximizing the size of $\Delta$ is motivated by the general absence of exact knowledge on $x(0)$ beyond the assumption that it lies within $\mathcal{X}_{\mathrm{safe}}$, 
	thus we seek to increase the validity of the controller we have designed over unknown initial conditions $x(0)$.  
	Once a set $\Delta$ is computed (regardless of whether it is maximized with respect to a parameter), 
	we have both synthesized a controller that meets performance specifications and obtained a guarantee on its safety given the set of initial states.  
				
	There are several ways to compute safety-invariant sets for dynamical systems; here we use an approach based on reachability analysis \cite{Oishi_CDC06}.  The safety-invariant set is the complement of the set of initial states that will eventually exit $\mathcal{X}_{\mathrm{safe}}$, and therefore we can compute the \emph{backwards reachable set} starting from all states outside of $\mathcal{X}_{\mathrm{safe}}$.  
	The set complement, $\mathcal{X}_{\mathrm{safe}}^c$, is propagated backwards in time through the closed loop dynamics \eqref{eq:dynamicsFB} to find the set of initial states starting inside of $\mathcal{X}_{\mathrm{safe}}$ that end in $\mathcal{X}_{\mathrm{safe}}^c$.  This can be done for a purely deterministic system (when the reference input $v$ in \eqref{eq:stateFB} is identically zero) or a non-deterministic system (reference input is allowed to vary).  
In the non-deterministic case, the safety-invariant set may encompass trajectories that remain within $\mathcal{X}_{\mathrm{safe}}$ for \emph{any possible} choice of $v$ (i.e. even ``adversarial'' inputs that try to drive $x(t)$ outside of $\mathcal{X}_{\mathrm{safe}}$), or for a \emph{single}, optimal input $v$, that acts to keep $x(t)$ within $\mathcal{X}_{\mathrm{safe}}$.  For instance, the reference input $v$ can be treated as the parametrization constant $\beta$ of the linearising controller $g(x,\beta)$, and the safety-invariant set is then maximized with respect to $v$ (while still satisfying the saturation constraints). 	For deterministic dynamics with $v=0$, we can still parametrize the controller when there are multiple possibilities to linearise the system (an instance is discussed in the case study).  Note that in this case we limit our focus to a \emph{constant} parameter $\beta$, whereas in the non-deterministic case $v$ is typically allowed to change with time, hence we consider signals $v(t)$.

	The backwards reachable set is computed using a Hamilton-Jacobi formulation (which allows us to minimize the safety-invariant set over $\beta(t)$, if desired) \cite{Mitchell2005_TAC}, and while difficult to solve exactly in practice, can be approximated using well-established discretization methods, such as those implemented in the Level Set Toolbox \cite{Mitchell2}.  Other computational approaches include viability kernel methods \cite{Maidens2013}, which rely on propagating certain types of shapes through linear dynamics to avoid discretization.  
				
	\medskip 				
				
	In this work, state $x$ is not available as input to function $g$, so we must compute $\Delta$ under the additional constraint that the control input is provided by $g(\hat{x})$, with $\hat{x}$ the output of an HGO.  
	The safety-invariant set computed under output feedback, rather than state feedback, will be denoted as $\tilde{\Delta}$. 
	While computing reachable (and hence safety-invariant) sets is well established given the feedback controller $g(x)$ and a possible reference signal $v$, 
	neither linearisation nor satisfaction of the control constraints are guaranteed under the output feedback controller $g(\hat{x})$, and the computation of $\tilde \Delta$ becomes more difficult.  First, we would need to keep track of both $x$ and $\hat{x}$, doubling the dimensionality of the system, which severely limits the applicability of discretization methods.  Second, we would no longer have linear closed-loop dynamics, and could not use the viability kernel methods that are suitable to higher dimensional systems so long as the dynamics are linear.
	
	It is therefore desirable to first design a controller under the assumption that the state is available for feedback, compute $\Delta$, and then \emph{formally relate} $\Delta$ to $\tilde{\Delta}$ under output feedback. In summary: 
	
	\smallskip
	
	\begin{prob}
		Given a full state feedback linearisable system \eqref{eq:dynamics} whose state is not fully available for control, 
		\begin{enumerate}
			\item
			compute a high-gain observer \eqref{eq:observer} that converges to the original dynamics \eqref{eq:dynamics} as quickly as possible; 
			\item
			compute a feedback linearising controller  $g(x)$ (or $g(x,\beta)$), and the corresponding safety-invariant set $\Delta$ (possibly maximizing $\Delta$ as a function of $\beta$), and quantitatively relate $\Delta$ to $\tilde \Delta$ under the condition that $g(x)$ is replaced by $g(\hat{x})$. \hfill $\square$
		\end{enumerate}		
	\end{prob}

	\smallskip

	\section{Main Results}
	
	In order to address Problem 1, we first provide a result on the convergence of trajectories under output and state feedback, which  
	generates an upper bound on the distance between closed-loop trajectories. We can therefore first compute $\Delta$, and use the bound between state and input trajectories to derive quantitative claims about $\tilde \Delta$.
	
	\subsection{Convergence of State- and Output-Feedback Trajectories}
	
	The trajectory $x(t)$ under output feedback recovers the desired trajectory $\overline{x}(t)$ under state feedback, namely the solution of $\dot{\overline{x}} = (A + BK)\ol{x} + Bv$, provided Assumptions \ref{ass3} and \ref{ass:xSafe} are satisfied, and $\hat{x}(0)\in\mathcal{X}_{\mathrm{safe}}$.  
	This known result \cite{Khalil_IJRNC14} is typically presented in the context of feedback stabilization, 
	and asymptotic guarantees on convergence are provided as a function of $\epsilon\rightarrow 0$ (the parameter of the HGO).   
	In this work, rather than only providing asymptotic results \cite{Khalil_TAC99}, 
	we discuss precise upper bounds on the distance between $x(t)$ and $\ol{x}(t)$, as a function of $\epsilon$.  
	Further, whereas \cite{Khalil_TAC99} shows that stability of the origin is preserved under output feedback, and gives an $O(\epsilon)$ bound on the distance between trajectories based on their convergence towards the origin, we show convergence of trajectories to each other as $\epsilon\rightarrow 0$ without the assumption that the control law is stabilizing.  Note that without stability assumptions, the following result is only valid for finite time horizons $T<\infty$ (we will later address the infinite horizon case).  
	
	\smallskip
	
	 \begin{thm}\label{thm:trajectoryConv} 
	 	Given a nonlinear system \eqref{eq:dynamics} satisfying Assumptions \ref{ass3}-\ref{ass:xSafe}, 
		a time horizon $T<\infty$, 
		and a desired bound $\xi>0$, 
		there exists a parameter $\epsilon$ for 
		a high-gain observer \eqref{eq:observer}, 		
		such that $\|x(t) - \ol{x}(t)\|_2\leq \xi$ for all $0\leq t \leq T$. 
		\hfill$\square$
	 \end{thm}

	\smallskip

	The proof follows directly from singular perturbation theory, and in particular what is known as Tikhanov's Theorem \cite{Wasow}.  
	We focus on characterising the bound $\xi$ exactly as a function of $\epsilon$, and show that $\xi$ can be made arbitrarily small as $\epsilon\rightarrow 0$.  For verification purposes, however, we are more interested in the value of $\xi$ for a given $\epsilon$, rather than the limiting behaviour as $\epsilon\rightarrow 0$, 
	since the earlier will establish a connection between $\Delta$ and $\tilde{\Delta}$.  
	Note that the restriction of $x(t)$ to $\mathcal{X}_{\mathrm{safe}}$ means that in practice we only require $\mathcal{A}=\mathcal{X}_{\mathrm{safe}}$, i.e. we do not require Lipschitz or boundedness conditions on \eqref{eq:dynamics} outside of $\mathcal{X}_{\mathrm{safe}}$.

	The relation to singular perturbation theory is seen by replacing the observer dynamics with a scaled version of the estimation error: 	
		\begin{equation}
		\label{eq:eta}
		\eta_i = \frac{x_i - \hat{x}_i}{\epsilon^{n-i}}. 
		\end{equation}
		We can then write $x - \hat{x} = D(\epsilon)\eta$, with $D(\epsilon)$ a diagonal matrix with entries $[\epsilon^{n-1},\dotsc,\epsilon, 1]$.  
		The combined state and observer dynamics can then be written as 
		\begin{equation}\label{eq:singPert}
		\begin{aligned}
		\dot{x} &= Ax + B[b(x) + a(x)g(x - D(\epsilon)\eta)]\\
		\epsilon\dot{\eta} &= \Lambda \eta + \epsilon B[b(x) + a(x)g(x - D(\epsilon)\eta) \\
		&\hspace{15mm}- b(\hat{x}) - a(\hat{x})g(x - D(\epsilon)\eta)], 
		\end{aligned}
		\end{equation}
		with 
		\begin{equation*}
		\Lambda = \begin{bmatrix} -\alpha_1 & 1 & 0 &\cdots & 0 \\ -\alpha_2 & 0 & 1 &\dotsc & 0 \\ \vdots & \vdots & \ddots & \ddots & \vdots 
		 \\ \vdots & \vdots & \cdots & \ddots & 1 \\ -\alpha_n & 0 & \cdots & 0 & 0 \end{bmatrix}. 
		\end{equation*}
		The system \eqref{eq:singPert} is a standard singularly perturbed system \cite{Wasow}.  
		We can think of the $x$ dynamics as the ``slow'' subsystem, 
		and the $\eta$ dynamics as the ``fast'' subsystem. 
		The design of the matrix $H(\epsilon)$ for the high-gain observer in \eqref{eq:observer} ensures that the matrix $\Lambda$ is stable.  
		If we scale time through the change of variable $\tau = \frac{t}{\epsilon}$, and then let $\epsilon=0$, the subsequent boundary-layer system
		\begin{equation}\label{eq:boundLayer}
		\frac{d\eta}{d\tau} = \Lambda \eta, 
		\end{equation}
		has an isolated, asymptotically stable root at $\eta=0$ \cite{Wasow}.  
		The reduced system that results from setting $\eta=0$ in \eqref{eq:singPert} is the system we would \emph{like} to have (system with state feedback designed to satisfy all the properties that we have specified), whose trajectory we denote by $\ol{x}(t)$ (cf. notation at the beginning of this section).  
		
		The motivation for studying the singularly perturbed problem is as follows.  
		First, the fast variable $\eta$ varies quickly, and is approximated by the boundary-layer system in \eqref{eq:boundLayer}.  
		Because of the difference in time scales, $x$ varies little in this period, and remains close to its initial condition $x(0)$.  
		Then, as $\eta$ has converged to its equilibrium point, 
		the slow variable takes over, and the behaviour of $x$ can be approximated by the reduced system $\ol{x}$ (obtained with $\eta = 0$).  
		We will formalize the amount that each variable can change during these time periods, in order to get bounds on the distance between $x(t)$ and $\ol{x}(t)$.
		
		Here we do not provide the full proof of Theorem \ref{thm:trajectoryConv}, which is quite lengthy and similar to that in \cite{Khalil_TAC99}, but outline the necessary steps. 
		We first make claims on the behaviour of $\eta$, and then describe the behaviour of $x$ as a function of $\eta$.   

		\begin{enumerate}
			\item 
			Because the boundary-layer system is stable, there exists a Lyapunov function $W(\eta) = \eta^TP\eta$, such that $\Lambda^TP + P \Lambda = -I$.  
			Further, there exists a constant $\rho>0$ such that $\Omega_{\rho}^{\eta} = \{\eta: W(\eta) \leq \rho \epsilon^2\}$ is a positively invariant set, 
			for any $\epsilon< \ol{\epsilon} = \frac{1}{4M_1\|P\|_2}$ ($M_1$ is the Lipschitz constant associated with the variable $x$ for $b(x) + a(x)u$, 
			to be further explained shortly).
			\item
			The variable $\eta$ enters $\Omega_{\rho}^{\eta}$ in finite time $T(\epsilon)$, with $\lim_{\epsilon\rightarrow 0}T(\epsilon)=0$.  
			\item
			Before $\eta$ enters $\Omega_{\rho}^{\eta}$, we can pick $ \epsilon = \epsilon_1$, $0< \epsilon_1<\bar \epsilon$, such that $\|x(t) - \ol{x}(t)\|_2\leq \xi$, for $0\leq t \leq T(\epsilon)$. 
			\item
			Once $\eta$ enters $\Omega_{\rho}^{\eta}$, we can choose $ \epsilon = \epsilon_2$, $0 < \epsilon_2 < \bar \epsilon$ such that $\|x(t) - \ol{x}(t)\|_2\leq \xi$.
		\end{enumerate}

	\smallskip 
	
	Without going into full details, we can elaborate upon the bounds on $\|x(t) - \ol{x}(t)\|_2$ in steps 3) and 4) above, as follows.  
	For $0 \leq t \leq T(\epsilon)$, 
	\begin{equation}\label{eq:boundTeps}
	\|x(t) - \ol{x}(t)\|_2\leq C_1\lambda_{\min}(P)\ln\left(\frac{k^2}{\epsilon^{2n}}\right)\epsilon,
	\end{equation}
	and for $T(\epsilon)\leq t \leq T$,
	\begin{multline}\label{eq:boundT}
	\|x(t) - \ol{x}(t)\|_2\leq \left(4C_1\lambda_{\min}(P)\ln\left(\frac{k^2}{\epsilon^{2n}}\right)C_2\right. \\
	+\frac{M_2\gamma\|D(\epsilon)\|_2}{1+M_1}C_2 - \left.\frac{M_2\gamma\|D(\epsilon)\|_2}{1+M_1} \right)\epsilon.
	\end{multline}

	The constants used in \eqref{eq:boundTeps} and \eqref{eq:boundT} are as follows. 
	The Lipschitz constant $M_1$ associated with the variable $x$ for $b(x) + a(x)u$ is guaranteed to exist by Assumption \ref{ass3}, since both $a(\cdot)$ and $b(\cdot)$ are Lipschitz continuous in $x$, and therefore $b(x) + a(x)u$ is Lipschitz in $x$. 
	Constant $M_2$ is defined as $\max_{x\in\mathcal{X}_{\mathrm{safe}}}|a(x)|$, 
	and $\gamma$ is the Lipschitz constant associated with $g(x)$ defined in \eqref{eq:stateFB}, which again exists by Assumption \ref{ass3}. The notation $\lambda_{\min}(P)$ refers to the smallest eigenvalue associated with matrix $P$.  
	We define constant $k$ as a bound on the initial distance between $x(0)$ and $\hat{x}(0)$, which we can guarantee exists if we set $\hat{x}(0)\in\mathcal{X}_{\mathrm{safe}}$ and impose Assumption \ref{ass:xSafe}. Given a neighbourhood $N_{r}(x(0))\subset \mathcal{X}_{\mathrm{safe}}$, based on Assumption \ref{ass3} we claim that 
	\begin{equation*}
	\|Ax + B[b(x)+a(x)u]\|_2\leq C_1, 
	\end{equation*}
	for all $x\in N_r(x(0))$.  We require $x(0)\in \mathcal{X}_{\mathrm{safe}}\backslash \partial \mathcal{X}_{\mathrm{safe}}$ so that the neighborhood $N_r(x(0))$ is contained in $\mathcal{X}_{\mathrm{safe}}$, and therefore Assumption \ref{ass3} applies. 
	Finally, $C_2=e^{(1+M_1)(T-T(\epsilon))}$.

	 For a given constant $\xi$, we can pick $\epsilon_1\leq\ol{\epsilon}$ small enough such that $\|x(t)-\ol{x}(t)\|_2\leq\xi$ for $0\leq t\leq T(\epsilon)$ using \eqref{eq:boundTeps}. We can also pick $\epsilon_2\leq \ol{\epsilon}$ such that $\|x(t) - \ol{x}(t)\|_2< \xi$ for $T(\epsilon)\leq t \leq T$ using \eqref{eq:boundT}.  
	 By taking $\epsilon = \min\{\epsilon_1, \epsilon_2\}$, we guarantee that $\|x(t) - \ol{x}(t)\|_2\leq \xi$ for all $0\leq t \leq T$.  The upper bound $\bar \epsilon$ is required to ensure the positive invariance of $\Omega_{\rho}^{\eta}$.  Notice that, as $\epsilon\rightarrow 0$, $\|x(t)-\ol{x}(t)\|_2\rightarrow 0$ uniformly over $0\leq t\leq T$.  Equations \eqref{eq:boundTeps} and \eqref{eq:boundT} therefore provide both a direct way of computing $\xi$ for a given $\epsilon$ (by taking the maximum of the right hand side of both equations) \emph{or} for verifying the maximum value $\epsilon$ can take to achieve a desired bound $\xi$.
	 
		\smallskip 	 
	 
	 Theorem \ref{thm:trajectoryConv} is not specific to feedback linearisable systems, i.e. $a(x)+b(x)u$ can be replaced by a more general Lispchitz continuous function $\phi(x,u)$.  
	 A possibly tighter bound is available for linearizable systems with control input \eqref{eq:stateFB}, because an explicit solution to \eqref{eq:dynamicsFB} is available of the form $x(t) = x(0)\exp\{(A + BK)t\}$, and therefore \eqref{eq:boundT} can be refined.
	
		\smallskip 
	 
	 The next Lemma considers the scenario when the linearising controller asymptotically stabilizes the system, and $g(x)$ is given by \eqref{eq:stateFB} with $v=0$ and $K$ chosen to stabilize $A+BK$.  
	 In this instance we can get a tighter bound on $\|x(t)-\ol{x}(t)\|_2$, 
		because $A+BK$ has negative eigenvalues, thus the quantity $\|\exp\{(A + BK)t\}\|$ does not increase with $t$.
	 	
		\smallskip 
		
	 	\begin{lem}\label{lem:tighterbound}
	 		For the feedback linearizable system in \eqref{eq:dynamics} satisfying Assumption \ref{ass3}, high-gain observer \eqref{eq:observer} with parameter $\epsilon$, controller $g(x) = \frac{-b(x) + Kx}{a(x)}$ with $K$ chosen such that $A+BK$ is stable, and time horizon $T$, 
	 		the upper bound on $\|x(t) - \bar x(t)\|_2$ is given by \eqref{eq:boundTeps} for $0\leq t \leq T(\epsilon)$, and by 
	 		\begin{multline*}
	 		\|x(t) -\ol{x}(t)\|_2 \leq	\left[\left(4C_1\lambda_{\min}(P)\ln\left(\frac{k^2}{\epsilon^{2n}}\right)\right)\hat C_2\right. \\
	 		\hspace{20 mm}+ \left. \vphantom{\left(4C_1\lambda_{\min}(P)\ln\left(\frac{k^2}{\epsilon^{2n}}\right)\right)}(T-T(\epsilon))M_2\gamma\|D(\epsilon)\|_2\right]\epsilon 
	 		\end{multline*}
	 		for $T(\epsilon) \leq t \leq T$.  \hfill $\square$
	 	\end{lem}
		
		\smallskip 		
	 		
	 		Notice that the matrix $A + BK$ is assumed to be diagonalisable, and if not it can be put into Jordan form.  
			Then $A+BK = G\Theta G^{-1}$, with $G$ the matrix of (generalised) eigenvectors, and $\Theta$ the diagonal or Jordan matrix containing the eigenvalues (with negative real part).  
			The constant $\hat C_2 = \|G\|\|G^{-1}\|\left\|e^{\Theta (T-T(\epsilon))}\right\|$, and all other constants coincide with those in Theorem \ref{thm:trajectoryConv}. 

	
%
	
%
	\subsection{Invariance Properties of Feedback Linearisable Systems}

	We can now use Theorem \ref{thm:trajectoryConv} (and, by extension, Lemma \ref{lem:tighterbound}) to show how to compute set $\tilde \Delta$ by computing the corresponding set $\Delta$ under state feedback.  
	We first provide a finite time horizon result, which does not require any stability assumptions on the controller $g(x)$. 
	We will denote, for a given parameter $\delta >0$, the restricted set 
	$(\mathcal{X}_{\mathrm{safe}}-\delta)=\{x:x\in\mathcal{X}_{\mathrm{safe}}\,\cap \|x-\tilde{x}\|_2\geq \delta,\,\forall\,\tilde{x}\in\partial\mathcal{X}_{\mathrm{safe}}\}$, 
	where $\partial\mathcal{X}_{\mathrm{safe}}$ is the boundary of $\mathcal{X}_{\mathrm{safe}}$. 
	
	\smallskip
	
		\begin{thm}\label{thm:Invariance}
			For the feedback linearisable system \eqref{eq:dynamics} satisfying Assumption \ref{ass3}, 
			the high-gain observer \eqref{eq:observer} with parameter $\epsilon$, 
			safe set $\mathcal{X}_{\mathrm{safe}}$ and control constraint $\mathcal{U}_{\mathrm{safe}}$, 
			and given the finite time horizon $T$, 
			 the safety invariant set $\Delta$, 
			computed using state feedback, $\mathcal{U}_{\mathrm{safe}}$ and  $(\mathcal{X}_{\mathrm{safe}}-\xi)$ (with $\xi$ the bound between trajectories $x(t)$ and $\bar x(t)$ provided in Theorem \ref{thm:trajectoryConv}),  
			is an underestimate of the safety invariant set $\tilde \Delta$ under output feedback for the original safe set $\mathcal{X}_{\mathrm{safe}}$.   \hfill $\square$
		\end{thm}

	\smallskip
		
	Note that for some choices of $\epsilon$, the associated $\xi$ may exceed the diameter of $\mathcal{X}_{\mathrm{safe}}$, in which case $\tilde \Delta$ will be the empty set.  We assume that $\epsilon$ can be chosen small enough, and hence $\xi$ is small enough, to render $\tilde \Delta$ nonempty (although an empty $\tilde \Delta$ is also informative).  From Theorem \ref{thm:trajectoryConv}, we know that for any $0< \epsilon < \bar \epsilon$, there exists a corresponding $\xi$ such that $\|x(t) - \ol{x}(t)\|_2\leq \xi$ for all $t\in[0,T]$.  
		Therefore, for a given $\epsilon>0$, if we can control $\ol{x}(t)$ to remain inside $(\mathcal{X}_{\mathrm{safe}} - \xi)$ for $t\in[0,T]$, we guarantee that $x(t)\in\mathcal{X}_{\mathrm{safe}}$.

	\smallskip	
	
	We can extend Theorem \ref{thm:Invariance} to the infinite horizon case, 
	provided that the state feedback control law $g(x)$ asymptotically stabilizes the origin, as in Lemma \ref{lem:tighterbound}.  
	We again employ the control structure in \eqref{eq:stateFB} with $v=0$, and  
 $K$  chosen such that $A+BK$ has eigenvalues with negative real parts. 
	
	\smallskip
	
	\begin{thm}\label{thm:Invariance2}
		For the feedback linearisable system \eqref{eq:dynamics}  satisfying Assumptions \ref{ass3} and \ref{ass:xSafe}, 
		the high-gain observer \eqref{eq:observer} with parameter $\epsilon$, 
		safe set $\mathcal{X}_{\mathrm{safe}}$, and control constraint $\mathcal{U}_{\mathrm{safe}}$, 
		 the infinite horizon safety-invariant set $\Delta$, 
		computed using state feedback and constraint sets $(\mathcal{X}_{\mathrm{safe}}-\xi)$ and $\mathcal{U}_{\mathrm{safe}}$ ($\xi$ is again the bound from Theorem \ref{thm:trajectoryConv}), is an underestimate of the safety invariant set $\tilde \Delta$ under  output feedback for the original constraint sets $\mathcal{X}_{\mathrm{safe}}$ and $\mathcal{U}_{\mathrm{safe}}$.  
		Further, the origin remains asymptotically stable under output feedback. \hfill $\square$
	\end{thm}

	\smallskip	

	We know from Theorem \ref{thm:Invariance} that, over a finite time horizon, we can underestimate the safety-invariant set $\tilde \Delta$ by assuming state feedback and using modified constraint sets $(\mathcal{X}_{\mathrm{safe}}-\xi)$ and $\mathcal{U}_{\mathrm{safe}}$.  
	We can extend Theorem \ref{thm:Invariance} to the infinite horizon by constructing a positively invariant set $\Omega_c^x\times\Omega_{\rho}^{\eta}$, with $\Omega_{\rho}^{\eta}$ the same as in Theorem \ref{thm:trajectoryConv}, and $\Omega_c^x \subset \Delta$, the safety-invariant set using an asymptotically stabilizing state feedback controller and constraint sets $\mathcal{X}_{\mathrm{safe}}$ and $\mathcal{U}_{\mathrm{safe}}$.  We only need $\Omega_c^x$ to be invariant while respecting the original constraint set $\mathcal{X}_{\mathrm{safe}}$ (if $\Omega_c^x\subset\Delta$, then $x(t)$ is guaranteed to not leave $\mathcal{X}_{\mathrm{safe}}$).   
	
	Under state feedback (where the control law renders the origin asymptotically stable), there exists a Lyapunov function $V(\bar x) = \bar{x}^T Q \bar{x} \leq -\|\bar{x}\|_2^2$ for $x\in\Delta$.  By definition, the set $\Omega_c^x = \{\bar x: V(\bar x)\leq c\} \subset \Delta$ is positively invariant.  For $\eta\in\Omega_{\rho}^{\eta}$, the set $\Omega_c^x$ remains positively invariant under output feedback.

	 We can show that for $t \geq T_1$, where 
	\begin{equation}\label{eq:T1}
	T_1 =  \frac{\lambda_{\mathrm{max}}(Q)}{2}\ln\left(\frac{\lambda_{\mathrm{max}}(Q)\|x(0)\|_2^2}{c}\right),
	\end{equation} 
	 $x(t)$ (using output feedback) will have reached $\Omega_c^x$.  Note that  we would like to make $c$ as large as possible, to reduce the time it takes for $x(t)$ to reach $\Omega_c^x$.
	 
	 Therefore, after time $T_1$, we can guarantee that $(x, \eta)\in \Omega_c^x\times\Omega_{\rho}^{\eta}$, with $\Omega_c^x\subset\Delta$, and that $(x,\eta)$ will remain inside $\Omega_c^x\times \Omega_{\rho}^{\eta}$ for all $t\geq T_1$.  Hence for all $t\geq T_1$, we are guaranteed that $x(t)$ does not violate the safety constraints, and $g(\hat{x})$ does not violate the control constraints.  Before time $T_1$, we do not have the guarantee on the state constraints, and apply Theorem \ref{thm:Invariance} over the finite time horizon $[0,T_1]$.  We compute $\xi$ as in Theorem \ref{thm:Invariance}, using \eqref{eq:boundTeps} and \eqref{eq:boundT} from Theorem \ref{thm:trajectoryConv}, and shrink $\mathcal{X}_{\mathrm{safe}}$ appropriately, then compute $\tilde\Delta\subset \Delta$, which is the infinite horizon safety-invariant set.
	 
	 In order to ensure the invariance of $\Omega_c^x$ and $\Omega_{\rho}^{\eta}$, we require an upper bound on the value that $\epsilon$ can take.  
	 From Theorem \ref{thm:trajectoryConv}, we know that $\epsilon < \bar \epsilon = \frac{1}{4M_1\|P\|}$ to ensure that $\Omega_{\rho}^{\eta}$ is invariant.  
	 To ensure that $\Omega_c^x$ is invariant and that all trajectories outside of $\Omega_c^x$ enter $\Omega_c^x$ in finite time requires $\epsilon < \min\{\epsilon_3, \epsilon_4\}$, with
	 \begin{equation*}
	 \epsilon_3 = \frac{\beta}{2\lambda_{\mathrm{max}}(Q)x_{\mathrm{max}}L\gamma}, 
	 \end{equation*}
	 and $\epsilon_4$ chosen such that
	 	\begin{equation*}\label{eq:c}
	 	c\geq 16\lambda_{\mathrm{max}}(Q)^3L^2\gamma^2\|D(\epsilon)\|_2^2\epsilon_2^2
	 	\end{equation*}
	 is satisfied, while at the same time $\Omega_c^x\subset \Delta$.  The constant $L$ is equal to $\max_{x\in\Omega_c^x}|a(x)|$, and $\beta = \min_{x\in\partial \Omega_c^x}\|x\|_2^2$.  Therefore we require $\epsilon < \hat{\epsilon} = \min\{\bar{\epsilon}, \epsilon_1, \epsilon_2\}$.
	 
	 The recovery of asymptotic stability of the origin can be shown in the same manner as in \cite{Khalil_TAC08}. 
	  
%
%
%

	\medskip

To summarize, we can compute a finite or infinite horizon safety-invariant set under output feedback with the following procedures.  
\subsubsection{Finite horizon case}
\begin{itemize}
	\item Compute $\xi$ as a function of $\epsilon$ and time horizon $T$ according to \eqref{eq:boundTeps} and \eqref{eq:boundT} from Theorem \ref{thm:trajectoryConv}.
	\item
	Compute $\tilde \Delta$ using standard reachability techniques for fully observable systems, with state constraints $(\mathcal{X}_{\mathrm{safe}} - \xi)$, control constraints $\mathcal{U}_{\mathrm{safe}}$, and time horizon $T$.
\end{itemize}
\subsubsection{Infinite horizon case}
\begin{itemize}
		\item
		Compute $\Delta$ using $\mathcal{X}_{\mathrm{safe}}$ and $\mathcal{U}_{\mathrm{safe}}$.
		\item
		Find the largest $c$ such that $\Omega_c^x\subset \Delta$.
		\item
		Compute $T_1$ according to \eqref{eq:T1}.
		\item 
		Compute $\xi$ as a function of $\epsilon$ and $T_1$ according to Theorem \ref{thm:trajectoryConv}.
		\item
		Compute $\tilde \Delta$ using standard reachability techniques for fully observable systems, with state constraints $(\mathcal{X}_{\mathrm{safe}} - \xi)$, control constraints $\mathcal{U}_{\mathrm{safe}}$, and time horizon $T_1$.
\end{itemize}

The obtained set $\tilde \Delta$ is the safety-invariant set for either the finite or infinite horizon using a controller designed for state feedback, but whose input is an estimate of the state produced by a high-gain observer.
A simple case study that demonstrates the outlined techniques for computing $\tilde \Delta$ is provided in an extended version of this paper, available at XXXXX.

	\section{Computational Example} 
	
	We consider a simple example to elucidate the synthesis of $g(x)$, the calculation of $\epsilon$ and $\delta$, and the construction of $\Delta$ and $\tilde\Delta$.  
	The known double integrator has linear dynamics
	\begin{equation}\label{eq:doubleIntDyn}
	\begin{bmatrix} \dot{x}_1 \\ \dot{x}_2 \end{bmatrix} = \begin{bmatrix} 0 & 1\\ 0 & 0\end{bmatrix}\begin{bmatrix} x_1 \\ x_2\end{bmatrix} + \begin{bmatrix} 0 \\ 1\end{bmatrix}u, 
	\end{equation}
	which are already in normal form. 
	The objective is to design a feedback 
	controller that stabilizes the system to the origin (in this simple case the system is already linear). 
	We set $u = g(x) = -\beta x_1 - \beta x_2$ to obtain the closed-loop dynamics 
	\begin{equation*}
	\dot{x} = \begin{bmatrix} 0 & 1 \\ -\beta & -\beta \end{bmatrix} x,
	\end{equation*}
	which is asymptotically stable with two complex roots under state feedback.  
	
	However, we also assume that only the first state $x_1$ is available, i.e. the output is $y = x_1$, and therefore implementing $g(x)$ exactly is impossible.  
	We therefore construct the following high-gain observer, in the form of \eqref{eq:observer}: 
	\begin{equation}\label{eq:doubleInt2}
	\dot{\hat{x}} = \begin{bmatrix} 0 & 1\\ -\beta & -\beta \end{bmatrix}\hat{x} +  \begin{bmatrix}
	\frac{\alpha_1}{\epsilon} \\ \frac{\alpha_2}{\epsilon^2}\end{bmatrix}[x_1 - \hat{x}_1]. 
	\end{equation}
	The constants $\alpha_1$ and $\alpha_2$ must be chosen such that $s^2 + \alpha_1 s + \alpha_2$ has negative roots:  
	they are set to $\alpha_1 = \alpha_2 = 4$, and the resulting polynomial has two roots equal to $-2$.  
	
	We further require that the state $x(t)$ does not leave the region $\mathcal{X}_{\mathrm{safe}} = \{x: -4\leq x_1\leq 4,\,-3\leq x_2\leq 3\}$, 
	and restrict the control input to the condition $|g(\hat{x})|\leq 1$.  
	In other words, we seek to stabilize the origin and respect the bounds $\mathcal{X}_{\mathrm{safe}}$ using a saturating control law of the form \eqref{eq:controller} with $u_{\max}=1$.  
	We would then like to find the safety-invariant set (as per Definition \ref{def:invariance}), 
	which is the set of all initial states $x(0)$ resulting in trajectories meeting all of the above requirements (stability, plus state and input constraints). 
	
	Following Theorem \ref{thm:trajectoryConv}, we let $\eta= [\frac{x_1 - \hat{x}_1}{\epsilon}, x_2 - \hat{x}_2]$.  The composed singularly perturbed system is
	\begin{align*}
	\dot{x} &= Ax + Bg(x - D(\epsilon)\eta)\\
	\epsilon \dot{\eta} &= \Lambda \eta,  
	\end{align*}
	with $\Lambda = [-4,\, 1; -4,\, 0]$.  
	Notice that in this example the term $\epsilon B[b(x) + a(x)g(x -D(\epsilon)\eta)) - b(\hat{x}) -a(\hat{x}) g(x - D(\epsilon)\eta))]$ vanishes because $b(x)=0$ and $a(x)=1$.

\begin{figure}
		\centering
		\includegraphics[width=.8\linewidth,keepaspectratio]{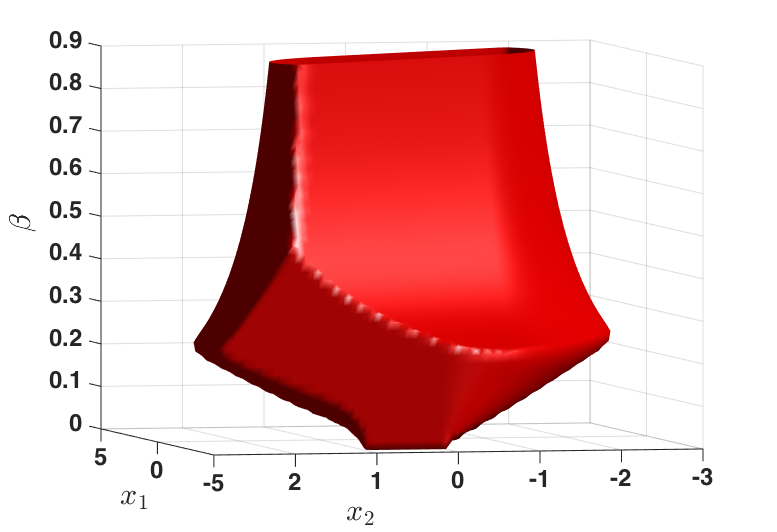}
		\caption{The safety invariant set, 
			computed using the Level Set Toolbox \cite{Mitchell2}, 
			using state feedback and fixed control law $g(x) = -\beta x_1 - \beta x_2$, for varying values of $\beta$.}
		\label{fig:stateFBbeta}
	\end{figure}

	Since our first objective was to stabilize \eqref{eq:doubleIntDyn} to drive trajectories to the origin, we do not want an additional reference input $v$.  We can, however, maximize the safety-invariant set by choosing $\beta$ appropriately.  We compute $\Delta$ for varying $\beta$ using a Hamilton-Jacobi formulation and the Level Set Toolbox, as outlined in \cite{Oishi_CDC06}. 
	Fig. \ref{fig:stateFBbeta} shows the safety-invariant set for varying $\beta$ (on the $z$-axis): 
	around $\beta = 0.2$ the safety-invariant set is the largest, and
	we therefore select the controller $g(x) = -0.2x_1 - 0.2x_2$.

\begin{figure}
		\centering
		\includegraphics[width=.8\linewidth,keepaspectratio]{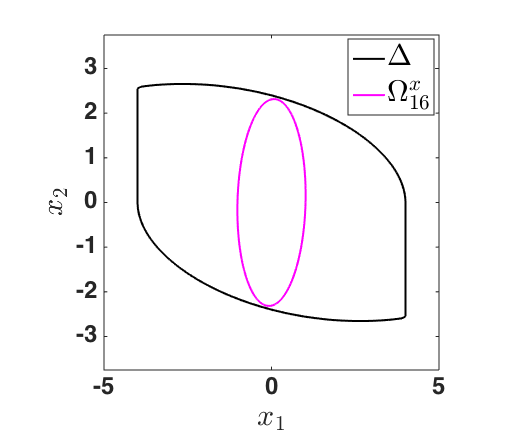}
		\caption{The safety-invariant set $\Delta$  using state feedback and the fixed control law $g(x) = -\beta x_1 - \beta x_2$, for $\beta=0.2$ (in black, external curve) and the maximal invariant set $\Omega_c^x = \{x:V(x)\leq c\}$ for $c=16$ (in purple, smaller ellipsoid).  }
		\label{fig:Omega_c}
	\end{figure}

	Under state feedback, the origin is asymptotically stable and there exists a Lyapunov function $V(x) = x^TPx$ satisfying $\dot{V}(x)\leq 0$. 
	Next, as described in Theorem \ref{thm:Invariance2}, we would like to find the largest set $\Omega_c^x = \{x: V(x)\leq c\}\subset \Delta$, with $\Delta$ the safety invariant set associated with $\beta = 0.2$ from Fig. \ref{fig:stateFBbeta}.  
	The set $\Omega_c^x$ with $c = 16$ is depicted inside $\Delta$ in Fig. \ref{fig:Omega_c}.  
	The constant $c$ was found manually by simply testing various possibilities and visualising them, 
	although a better way of doing this should be found for higher dimensional model instances.

\begin{figure}
		\centering
		\includegraphics[width=.8\linewidth,keepaspectratio]{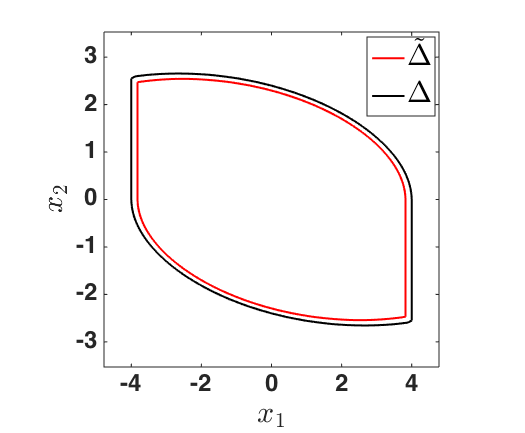}
		\caption{The safety-invariant set using state feedback and $\mathcal{X}_{\mathrm{safe}}$ (external black line) versus the output feedback safety invariant set, computed using $(\mathcal{X}_{\mathrm{safe}} - \delta)$ (internal red curve), for $\epsilon = 0.01$ and $\delta = 0.1768$.}
		\label{fig:safetyOFB}
	\end{figure}

	We  compute $T_1$ and $T(\epsilon)$, and then finally $\delta$.  For $\epsilon=0.01$, $T(\epsilon) = 0.0608$ seconds, $T_1 = 24.74$ seconds, and $\delta = 0.1768$.  
	If we instead picked $\epsilon = 0.001$, we could reduce $\delta$ to $0.022$.  
	As expected, as $\epsilon$ is made smaller, $\delta$ approaches 0, and we can essentially recover the state feedback safety-invariant set. 
	Restricting $\epsilon$ to 0.01, the safety-invariant set using $(\mathcal{X}_{\mathrm{safe}} - \delta)$ is depicted in Fig. \ref{fig:safetyOFB}, with the original safety-invariant set included for comparison.  
	
\begin{figure}
		\centering
		\includegraphics[width=.8\linewidth,keepaspectratio]{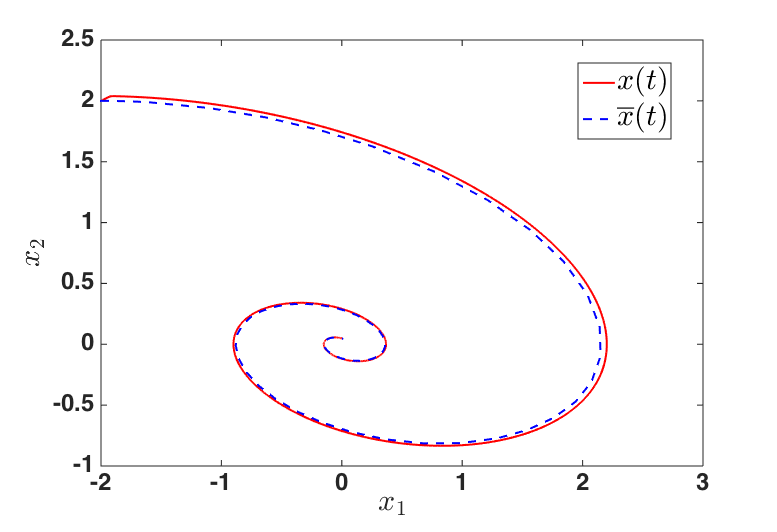}
		\caption{Sample trajectories initialised at $(-2,2)$ under state feedback ($\ol{x}(t)$) (blue, dashed) versus output feedback ($x(t)$) (red).  
			Note the initial trajectories difference over the output feedback curve, which is quickly corrected at time $T(\epsilon)$.} 
		\label{fig:trajectoryCompare}
	\end{figure}

\begin{figure}
		\centering
		\includegraphics[width=.8\linewidth,keepaspectratio]{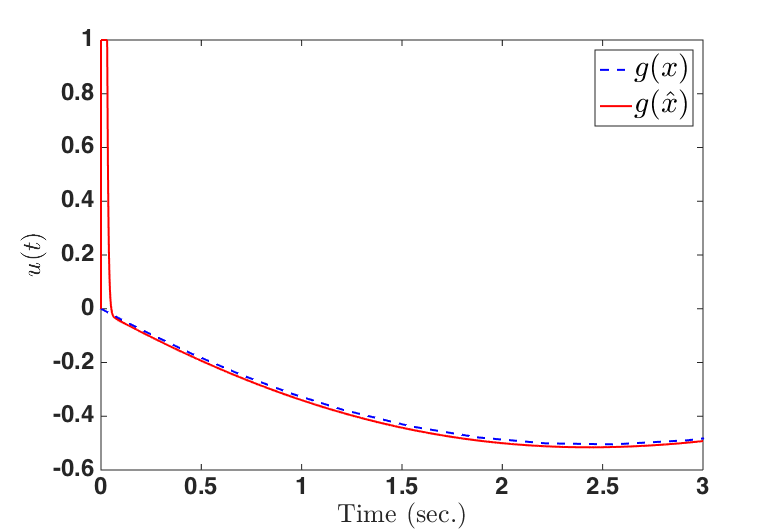}
		\caption{Control inputs generated for the trajectories in Fig. \ref{fig:trajectoryCompare}.  The state feedback control inputs are in blue (dashed) and the output feedback inputs are in red (solid line).  Note the initial saturation of the control inputs for the output feedback trajectory, which corresponds to the initial difference noticeable in Fig. \ref{fig:trajectoryCompare}.  After the initial saturation, however, the control inputs quickly almost exactly coincide.}
		\label{fig:controller}
	\end{figure}

	A sample trajectory of $\ol{x}(t)$ under state feedback is compared to the trajectory $x(t)$ under output feedback with $\epsilon=0.01$, starting from $x(0) = (-2,-2)$, in Fig. \ref{fig:trajectoryCompare}. Note the slight difference with $x(t)$ when the trajectory initializes, which is caused by the initial transient period before $T(\epsilon)$, when the output feedback controller is saturated and the state feedback controller is not.  The saturation can be seen in Fig. \ref{fig:controller}, which compares the state feedback control input to the output feedback control input.
	
	This example highlights that not only can we use the presented approach for verification of output feedback systems, 
	but also for the synthesis of controllers that satisfy stability and safety specifications under output feedback, 
	which is a consequence of being able to separate state estimation and control using the high-gain observer.  
	We can construct a controller that maximises the safety-invariant set $\Delta$ under state feedback, 
	apply the same controller to the system with output feedback, 
	and simultaneously produce the maximal safety-invariant set under output feedback.  The one drawback of course is that we do not know $x(0)$ exactly, and do not necessarily know if it falls within $\tilde \Delta$.  Hence the need to maximize $\tilde \Delta$, and also to exploit any knowledge we may have about $x(0)$ to try to ensure $\tilde \Delta$ contains the region in which $x(0)$ lies.

	\section{Conclusions}

	The use of a high-gain observer design allows us, for a specific class of feedback linearisable models, 
	a) to design controllers to satisfy certain specifications (in this work, safety) as if the full state were available for feedback, and 
	b) to derive a bound on the distance between the trajectories under state and output feedback, as a function of the observer parameter $\epsilon$.  
	Given this quantitative bound, if we can select $\epsilon$ as small as needed, we can completely recover whatever properties are satisfied under full state feedback. 
	
	\smallskip
	
	Possible extensions include feedback linearisable systems with zero dynamics and model uncertainty, which are discussed in \cite{Khalil_TAC08} and are aligned with the approach discussed in this work.  
	On the other hand, convergence of $\hat{x}$ to $x$ using a high-gain observer in the presence of output disturbances (e.g., $y = x_1 + w)$ unfortunately does not hold.  
	However, we may still be able to obtain a bound on the distance between trajectories, even if we cannot claim that the bound converges to zero with time.  
	Another straightforward extension is to consider more complex temporal properties, such as reach-avoid or other LTL specifications.  
	The difficulty of dealing with more general LTL specifications is in the construction of the state feedback controller, which typically is done by resorting to formal (finite-state) abstractions \cite{TabuadaBook, Zamani_TAC14}, 
	possibly coupled downstream with the design of an incrementally stabilizing controller, and which may violate the Lipschitz continuity assumptions required for the feedback controller in this work. 
	
		\bibliographystyle{IEEEtran}

\end{document}